\documentclass[12pt]{article}

\usepackage[dvips]{graphics}
\usepackage{booktabs,lscape,amssymb,amsmath}

\newcommand{\pal}{\partial}

\newcommand{\be}{\begin{equation}}
\newcommand{\ee}{\end{equation}}
\newcommand{\bea}{\begin{eqnarray}}
\newcommand{\eea}{\end{eqnarray}}
\newcommand{\bi}{\begin{itemize}}
\newcommand{\ei}{\end{itemize}}
\newcommand{\bay}{\begin{array}}
\newcommand{\eay}{\end{array}}

\newcommand{\tr}{{\rm tr}}

\newcommand{\non}{\nonumber}
\newcommand{\Dw}{D_{\rm w}}

\newcommand{\Dov}{D_{\rm N}}
\newcommand{\DWF}{D_{\rm DWF}}
\newcommand{\PV}{D_{\rm PV}}
\newcommand{\Dq}{D_{\rm q}}
\newcommand{\Deff}{D_{\rm eff}}

\newcommand{\pbf}{{\bf p}}

\newcommand{\xbf}{{\bf x}}
\newcommand{\ybf}{{\bf y}}
\newcommand{\zbf}{{\bf z}}
\newcommand{\ubf}{{\bf u}}
\newcommand{\vbf}{{\bf v}}

\title{
\begin{flushright}
{\normalsize CU-TP-1191}\\
\end{flushright}
A formulation of domain-wall fermions\\
in the Schr\"odinger functional
}

\author{Shinji Takeda\footnote{Present address:
Institute for Theoretical Physics, Kanazawa University,
Kanazawa 920-1192, Japan.
E-mail address: takeda@hep.s.kanazawa-u.ac.jp
}\\
Physics Department, Columbia University,\\
New York, NY 10027 USA
}

\begin{document}

\maketitle
\abstract{
We present a formulation of domain-wall fermions
in the Schr\"odinger functional by following a
universality argument.
To examine the formulation, we numerically investigate
the spectrum of the free operator and perform a one-loop analysis
to confirm universality and renormalizability.
We also study the breaking of the Ginsparg-Wilson relation
to understand the structure of chiral symmetry breaking
from two sources: The bulk and boundary.
Furthermore, we discuss the lattice artifacts of the step scaling function
by comparing with other fermion discretizations.
}



\section{Introduction}
\label{sec:introduction}

In the study of CP violation by CKM unitary triangle analysis,
hadron matrix elements of four-fermion operators,
such as $B_{\rm K}$, play a vital role.
Accurate calculations of this quantity from first principles
are an important task for the lattice QCD community.
In such calculations, having chiral symmetry
is crucial to avoid an operator mixing problem
which causes uncontrollable systematic errors.
Although lattice chiral fermions
 \cite{Kaplan:1992bt,Shamir:1993zy,Neuberger:1997fp}
are a clean formulation,
they require enormous computing power to perform dynamical simulations.
In comparison, ordinary fermion formulations, like Wilson type fermions
and starggared fermions are relatively cheap.
Nowadays, however, thanks to the development of computer architecture
and algorithms, dynamical simulations with
lattice chiral fermions have become feasible
even for three flavors \cite{Noaki:2008gx}.
In particular,
the RBC/UKQCD collaboration \cite{Allton:2008pn} is currently using
domain-wall fermions (DWFs) to compute $B_{\rm K}$.
In the course of their computation,
there are many sources of systematic errors which one has to control.
Among them, the non-perturbative renormalization (NPR)
could be serious.
At the moment, the collaboration has been using conventional schemes, such as,
the RI/MOM scheme and its variants \cite{Aoki:2007xm,Sturm:2009kb}.
However, these schemes potentially contain ``large scale
problem" which requires a quite large lattice volume.
To avoid such difficulties,
a new scheme was invented, known as the Schr\"odinger functional
(SF) scheme \cite{Luscher:1992an}.
This scheme provides a reliable way of estimating errors in the NPR.
If one wants to use this scheme for the renormalization
of $B_K$ given by the RBC collaboration,
first of all, one has to formulate DWF in the SF setup.
This is the purpose of this paper.

While chiral fermions are useful
for computing the bare $B_{\rm K}$
to avoid the mixing problem, a formulation for such fermions
in the SF setup was a non-trivial task
because SF boundary conditions break chiral symmetry explicitly.
We will address this issue in the next section.
However, Taniguchi \cite{Taniguchi:2004gf} made the first attempt to formulate overlap fermions
by using an orbifolding technique.
Subsequently he provided a formulation for domain-wall fermions
and then he and his collaborators \cite{Nakamura:2008xz} calculated
a renormalized $B_{\rm K}$ in quenched QCD.
Sint \cite{Sint:2010eh} developed such
techniques by combining with a flavor twisting trick.
However, these orbifolding formulations are constrained by
the requirement that the number of flavors be even.
Thus, apparently such formulations
are incompatible with current trends toward dynamical three flavor simulations.
To overcome this difficulty, L\"uscher \cite{Luscher:2006df} gave
a completely different approach
relying on a universality argument,
dimensional power counting and symmetry considerations.
Some perturbative calculations were performed in Ref.~\cite{Takeda:2007ga}.
A crucial property of this formulation is that
there is no restriction on the number of flavors.
Since only overlap fermions were considered in Ref.~\cite{Luscher:2006df},
our main purpose here is to formulate
the other chiral fermions, namely, domain-wall fermions.

The rest of the paper is organized as follows.
Section \ref{sec:formulation} gives the formulation of domain-wall
fermions in the SF setup, after a brief
review of the universality argument.
We present several pieces of numerical evidence
in Section \ref{sec:spectrum} and \ref{sec:oneloop}
to show that our formulation is working properly.
We also discuss the lattice artifacts
for the step scaling function in Section \ref{sec:SSF}.
In the last section, we conclude by giving some remarks and outlook.

\section{Formulation}
\label{sec:formulation}
In the following, we assume that the reader
is familiar with the SF in QCD \cite{Luscher:1992an,Sint:1993un}.
After giving a brief reminder of the universality argument,
we give a formulation for DWF
and finally check the chiral symmetry breaking structure numerically.

\subsection{Universality argument}

In the massless continuum theory,
the Dirac operator $D$ satisfies the anti-commutation relation with
$\gamma_5$
\be
\gamma_5D+D\gamma_5=0.
\label{eqn:Dg5}
\ee
The above is true even in the SF setup, although the boundary conditions,
\bea
P_+ \psi(x)=0 &\mbox{ at }& x_0=0,
\label{eqn:SFBC0}
\\
P_- \psi(x)=0 &\mbox{ at }& x_0=T,
\label{eqn:SFBCT}
\eea
with $P_\pm=(1\pm\gamma_0)/2$,
break chiral symmetry explicitly.
Eq.(\ref{eqn:Dg5}) means that the operator itself
does not know about boundary conditions.
In the continuum theory, information such as boundary conditions
is embedded in the Hilbert space.
In fact, the corresponding propagator,
which is a solution of the inhomogeneous equation,
\be
DS(x,y)=\delta(x-y),
\ee
fails to satisfy the anti-commutation relation.
Instead, it follows
\bea
\lefteqn{\gamma_5S(x,y)+S(x,y)\gamma_5=}\non\\
&&
\int_{z_0=0}d^3\zbf S(x,z)\gamma_5 P_- S(z,y)
+
\int_{z_0=T}d^3\zbf S(x,z)\gamma_5 P_+ S(z,y).
\label{eqn:Sg5}
\eea
This can be derived by using partial integration
on the SF manifold which has two boundaries
at time slice $x_0=0$ and $T$.
The non-vanishing right-hand side in eq.(\ref{eqn:Sg5})
shows an explicit chiral symmetry breaking.
Since such a breaking term is supported
only on the time boundaries,
the chiral symmetry is preserved in a bulk.

If someone naively tries to formulate
chiral fermions on the lattice,
one may define an overlap operator, for example,
with the Wilson kernel in the SF setup \cite{Sint:1993un}.
However such an operator immediately satisfies
the Ginsberg-Wilson relation and thus cannot reproduce
eq.(\ref{eqn:Sg5}) in the continuum limit.
This indicates that such naive formulation
does not work and furthermore may
belong to another boundary universality class
which is not what we want.
In this way, it is a non-trivial task to
formulate chiral fermions in the SF setup.

Some years ago, L\"uscher \cite{Luscher:2006df} proposed a clever way
to overcome this situation.
First, consider the relation for the propagator
in eq.(\ref{eqn:Sg5}).
This indicates that the GW relation has to be modified
by boundary effects.
Thus one has to find a modified overlap operator
which breaks the GW relation near the time boundaries and correctly
reproduces eq.(\ref{eqn:Sg5}) in the continuum limit.
Actually, finding such a modified operator is not so hard.
However, a new question naturally arising
is how the SF boundary conditions emerge.
For the Wilson fermion case \cite{Sint:1993un},
because there is a transfer matrix,
it is natural for fermion fields
to follow the SF boundary conditions.
However for chiral fermions,
there is no such transfer matrix
which can be defined from nearest neighbor interaction
in the time direction.
Therefore it is not an easy task.

L\"uscher \cite{Luscher:2006df} gave another point of view
to see how fields respect the boundary condition.
In the quantum field theory,
the correlation function can tell you
what kinds of boundary conditions are imposed.
As an example, let us see
how the boundary conditions emerge for Wilson fermions
whose action is given by
\bea
S_{\rm w}
&=&
\sum_x\bar\psi(x)
\Dw(m)
\psi(x),
\\
\Dw(m)
&=&
\frac{1}{2}
\left[
\sum_\mu (\nabla_\mu+\nabla^{\ast}_\mu)\gamma_\mu
-a\sum_\mu \nabla_\mu^{\ast}\nabla_\mu
\right]+m,
\eea
where $\nabla_\mu$ and $\nabla_\mu^{\ast}$ are
forward and backward covariant difference operators respectively,
\bea
\nabla_\mu\psi(x)
&=&
\frac{1}{a}
\left[
U(x,\mu)\psi(x+a\hat\mu)-\psi(x)
\right],
\\
\nabla_\mu^{\ast}\psi(x)
&=&
\frac{1}{a}
\left[
\psi(x)-U(x-a\hat\mu,\mu)^{-1}\psi(x-a\hat\mu)
\right].
\eea
In the SF setup, the sum over $x$ in the action is a little bit subtle.
We assume that the dynamical fields are 
$\psi(x)$ with $a\le x_0 \le T-a$ and
the fields $\psi(x)$ with $x_0 \le 0$ and $T\le x_0$ are set to zero.
For this setup,
the propagator may be defined by
\bea
\langle\eta(x)\bar\psi(y)\rangle&=&a^{-4}\delta_{x,y},
\label{eqn:etapsi}
\\
\eta(x)&=&\frac{\delta S_{\rm w}}{\delta\bar\psi(x)}.
\label{eqn:etaS}
\eea
For $2a \le x_0 \le T-2a$, eq.(\ref{eqn:etaS}) turns out to be
\be
\eta(x)=\Dw(m)\psi(x).
\ee
On the other hand, at $x_0=a$, we obtain
\bea
\eta(x)&=&
\frac{1}{a}P_+\psi(x)-\nabla_0P_-\psi(x)
\non\\
&+&
\frac{1}{2}
\left[
 \sum_k (\nabla_k+\nabla^{\ast}_k)\gamma_k
-a\sum_k \nabla_k^{\ast}\nabla_k
\right]
\psi(x)
+m\psi(x).
\label{eqn:eta}
\eea
By substituting eq.(\ref{eqn:eta}) into eq.(\ref{eqn:etapsi}) with $x\neq y$
we obtain
\be
\frac{1}{a}P_+\langle\psi(x)\bar\psi(y)\rangle|_{x_0=a}
-
\nabla_0P_-\langle\psi(x)\bar\psi(y)\rangle|_{x_0=a}
+...
=0.
\ee
In the continuum limit, the first term is dominant
\be
\frac{1}{a}P_+\langle\psi(x)\bar\psi(y)\rangle|_{x_0=0}=0.
\ee
This shows that in the naive continuum limit,
the Dirichlet type boundary condition ($P_+\psi|_{x_0=0}=0$)
is stable against the Neumann one ($\nabla_0P_-\psi|_{x_0=0}=0$), and in the end
the SF boundary conditions in eq.(\ref{eqn:SFBC0}) emerge.
It is plausible that similar things happen also for the chiral fermions case,
as long as the locality and symmetry
are kept in a proper way,
although we expect that the coefficient of the lowest dimensional operators
($\frac{1}{a}P_+\psi$)
may be different from the above case,
and more higher dimensional terms may appear in eq.(\ref{eqn:eta}).
The important point here is that
continuum SF boundary conditions emerge dynamically
in the continuum limit of the correlation function.
This boundary condition is natural
and automatically guaranteed to emerge
from the dimensional order counting argument.
Therefore, when we construct chiral fermions in the SF,
we only have to prepare a modified operator
by introducing an additional term
which breaks the chiral symmetry near the time boundaries.
Once this is fulfilled, then
such an operator automatically
turns out to be the desired one in the continuum limit without fine tuning.
A final important note is that the form of the boundary term is irrelevant
as long as it will go into a preferred boundary universality class.
Therefore, there is a large amount of freedom when choosing boundary terms
and one can use this freedom for practical purposes.

Following these guiding principles,
L\"uscher \cite{Luscher:2006df} proposed the operator:
\bea
\bar{a}\Dov&=&1-\frac{1}{2}(U+\tilde U),
\\
U&=&A(A^{\dag}A+caP)^{-1/2},\hspace{7mm}\tilde U = \gamma_5 U^{\dag} \gamma_5,
\\
A&=&1+s-a\Dw(0),\hspace{7mm}\bar{a}=a/(1+s),
\label{eqn:overlap}
\eea
with the parameter in the range $|s|<1$.
$\Dw(0)$ is the massless Wilson operator in the SF.
The key point here is the presence of the $P$
term in the inverse square root which is given by
\be
aP(x,y)
=
\delta_{\xbf,\ybf}\delta_{x_0,y_0}(\delta_{x_0,a}P_-+\delta_{x_0,T-a}P_+).
\ee
Note that this term is supported near the time boundaries
and thus called a boundary operator.
The presence of this term breaks the GW relation explicitly
and the breaking is given by
\bea
\Delta_{\rm B}=\gamma_5\Dov+\Dov\gamma_5-\bar{a}\Dov\gamma_5\Dov.
\eea
It was shown in ref.~\cite{Luscher:2006df} that this term is local and
supported in the vicinity of the boundaries up to the
exponentially small tails.

Although this operator breaks chiral symmetry explicitly,
other symmetries (the discrete rotational symmetries, $C$, $P$ and $T$,
flavor symmetry and so on)
have to be maintained
since the boundary conditions in eq.(\ref{eqn:SFBC0},\ref{eqn:SFBCT}) are
invariant under these symmetries.
In addition, this operator has $\gamma_5$-Hermiticity.
In this way, the universality formulation
can avoid breaking important symmetries, such as the flavor symmetry.
This is a distinctive feature of this formulation
compared with the orbifolding technique,
where flavor symmetries cannot be maintained
or, there is a constraint on the number of flavors.

Before leaving this subsection,
let us summarized the guiding principles
of formulating chiral fermions in the SF setup.
What we learned from this construction
is that, for an original chiral fermion operator,
one has to introduce an additional term
to break the chiral symmetry and
then demand that
such breaking only appears near the time boundaries.
Furthermore, one must
maintain important symmetries as well as $\gamma_5$-Hermiticity.
Once these conditions are fulfilled,
it is automatically guaranteed that the such a lattice operator
will correctly reproduce the continuum results
according to the universality argument.

\subsection{Formulation of domain-wall fermions }
Let us apply the guiding principles given in the previous subsection
to domain-wall fermions.
We propose a massless\footnote{The mass term can be introduced
in the usual way, namely
$a^4m_{\rm f}\sum_{\xbf}\sum_{x_0=a}^{T-a}[\bar\psi(x,1)P_R\psi(x,L_s)+\bar\psi(x,L_s)P_L\psi(x,1)]$.}
domain-wall fermion action
\be
S=
a^4\sum_{x,x^\prime}\sum^{L_s}_{s,s^\prime=1}
\bar\psi(x,s)(\DWF)_{xs,x^\prime s^\prime}\psi(x^\prime,s^\prime),
\ee
where a massless operator with $L_s=6$
for example\footnote{We restrict ourselves to an even number of $L_s$,
which is the case usually implemented.}
in four dimensional block form is given by
\be
a\DWF=
\left[
\begin{array}{cccccc}
a\tilde{D}_{\rm w}&-P_L&0&0&0&cB\\
-P_R&a\tilde{D}_{\rm w}&-P_L&0&cB&0\\
0&-P_R&a\tilde{D}_{\rm w}&-P_L+cB&0&0\\
0&0&-P_R-cB&a\tilde{D}_{\rm w}&-P_L&0\\
0&-cB&0&-P_R&a\tilde{D}_{\rm w}&-P_L\\
-cB&0&0&0&-P_R&a\tilde{D}_{\rm w}\\
\end{array}
\right],
\label{eqn:DWF}
\ee
with the chiral projections,
\be
P_{R/L}=(1\pm\gamma_5)/2.
\ee
We also assume that the dynamical fields
are $\psi(x,s)$ with $a\le x_0 \le T-a$.
The block elements in eq.(\ref{eqn:DWF}) are four dimensional operators
and $a\tilde\Dw$ is given by
\be
a\tilde{D}_{\rm w}=a\Dw(-m_5)+1.
\ee
The domain-wall height parameter usually takes a value in a range $0<am_5<2$.

An important ingredient here is the presence
of $B$ in eq.(\ref{eqn:DWF}) terms in the cross diagonal elements.
The reason for this $s$-dependence
is to break the chiral symmetry
in a similar way to the usual mass term \cite{Christ:2004gc}.
As mentioned before, such chiral symmetry breaking should be
present only near time boundaries, therefore, we chose
the $B$ term as
\be
B(x,y)=\delta_{\xbf,\ybf}\delta_{x_0,y_0}\gamma_5
(\delta_{x_0,a}P_-+\delta_{x_0,T-a}P_+),
\label{eqn:Bpm}
\ee
which is supported near the boundaries.
In this way, the time dependence is fixed.
The spinor structure ($\gamma_5 P_\pm$) is determined by imposing
the discrete symmetries C, P and T and $\Gamma_5$-Hermiticity.
These requirements are not so strong to determine
the spinor structure completely and therefore there is some freedom.
The structure proposed here
is only one of many solutions.
Actually, we examined several choices
of the spinor structure in the boundary term
and confirmed numerically the universal results
in the continuum limit for the lowest eigenvalue.
In the following,
we take this boundary term in eq.(\ref{eqn:Bpm}).

The boundary coefficient $c$ is supposed to be non-zero
to correctly reproduce the continuum theory as we will see in
Section \ref{sec:spectrum}.
It also plays an important role to cancel
boundary O($a$) cutoff effects and has
a perturbative expansion
\be
c=c^{(0)}+c^{(1)}g_0^2 + O(g_0^4).
\ee
In the same way as in \cite{Takeda:2007ga},
we tune the first coefficient $c^{(0)}$
as a function of the domain-wall hight $am_5$,
\bea
c^{(0)}&=&0.5089-0.0067(am_5-1)+0.0488(am_5-1)^2
\non\\
&&
-0.0216(am_5-1)^3+0.0673(am_5-1)^4.
\label{eqn:ctreeL}
\eea
This is valid in the region where
bulk O($a$) can be neglected
\footnote{
Actually, we observe that
O($a$) improvement program does not work for small $L_s$
and values of $am_5$ which are far from $1$.
For example O($a$) terms in $f_A$ ad $f_P$
(defined in appendix \ref{subsec:operators})
at tree level
do not vanish simultaneously with the same value of $c^{(0)}$.
},that is, for sufficient large $L_s$.
In the process of this determination, one needs to define the operators
of the axial vector current and pseudo scalar density.
We give their definition together
with that of the conserved axial current in appendix \ref{subsec:operators}.

\subsection{Structure of chiral symmetry breaking at tree level}
\begin{figure}[!t]
\begin{center}
  \begin{tabular}{ccc}
  \hspace{-30mm}
  \scalebox{1.2}{\includegraphics{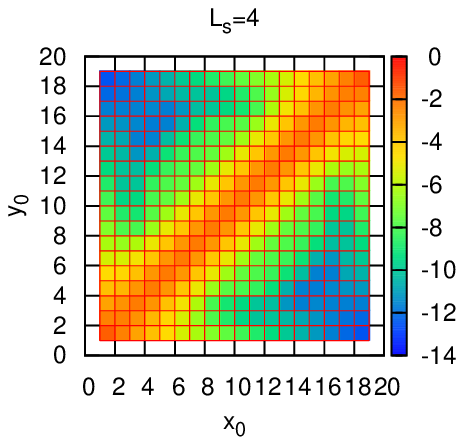}}&
  \hspace{-40mm}
  \scalebox{1.2}{\includegraphics{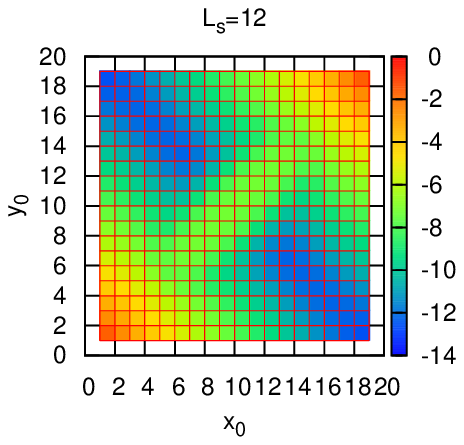}}&
  \hspace{-40mm}
  \scalebox{1.2}{\includegraphics{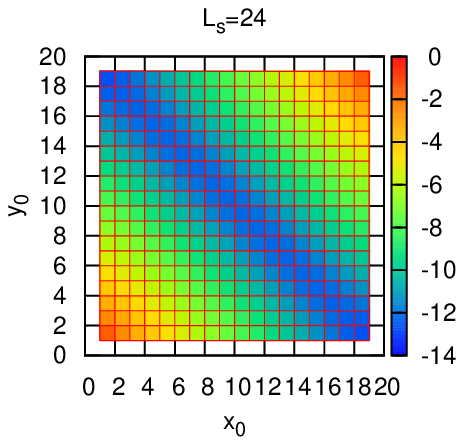}}\\
  \end{tabular}
\caption{
The color ranging $[-14,0]$ corresponds to the value of
$\ln||\Delta^{(L_s)}(x_0,y_0)||_{\rm spin}$
for the zero spatial momentum configuration with the parameters
$T/a=20$,
$m_{\rm f}=0$, 
$am_5=1$ and $c=1$.
\label{fig:GWbreaking}}
\end{center}
\end{figure}

In this subsection, 
let us check an important properties of the operator defined
in the previous section.
A reader may worry that even though the additional
boundary term is localized to time boundary,
after integrating over the fifth dimensional
degree of freedom such breaking effects may leak into the 4-dim bulk and
ruin the bulk chiral symmetry.
To settle this question, we numerically investigate the
structure of the chiral symmetry breaking
by looking at the breaking of the GW relation
\be
\Delta^{(L_s)}
=
\gamma_5\Deff^{(L_s)}+\Deff^{(L_s)}\gamma_5
-2a\Deff^{(L_s)}\gamma_5\Deff^{(L_s)},
\label{eqn:DeltaLs}
\ee
with the effective four dimensional operator
\cite{Kikukawa:1999sy,Kikukawa:1999dk}
\be
\det\Deff^{(L_s)}= \det[\DWF/D_{\rm PV}].
\ee
The Pauli-Villars (PV) operator is defined as
the massive DWF operator with $am_{\rm f}=-1$.
To obtain the effective operator,
first of all, we have to define physical quark fields
\bea
q(x)&=&P_L\psi(x,1)+P_R\psi(x,L_s),
\label{eqn:q}
\\
\bar{q}(x)&=&\bar{\psi}(x,1)P_R+\bar{\psi}(x,L_s)P_L.
\label{eqn:barq}
\eea
In terms of the propagator of domain-wall fermions defined from
\be
\DWF S_{\rm DWF}(x,y;s,t)=a^{-4}\delta_{x,y}\delta_{s,t},
\ee
that of the physical field is expressed
\bea
\left[q(x)\bar{q}(y)\right]_{\rm F} 
&=&
S_{\rm q}(x,y) 
\label{eqn:Sq}
\non\\
&=&
 P_LS_{\rm DWF}(x,y;1,L_s)P_L
+P_LS_{\rm DWF}(x,y;1,  1)P_R
\non\\
&+&
 P_RS_{\rm DWF}(x,y;L_s,L_s)P_L
+P_RS_{\rm DWF}(x,y;L_s,  1)P_R.
\eea
In terms of $S_{\rm q}$
the effective operator is given by
\be
a\Deff^{(L_s)}=(1+a^3S_{\rm q})^{-1}.
\ee

In the SF setup, $\Delta^{(L_s)}$ in eq.(\ref{eqn:DeltaLs})
contains not only the bulk chiral symmetry breaking
but also the boundary breaking.
The former is supposed to be removed by taking $L_s$ to infinity.
In such limit, boundary breaking effects remain
and they are expected to be localized near time boundaries.
To see this situation, we numerically compute $\Delta^{(L_s)}$
for a free operator.
In the free case, 
we can perform the Fourier transformation for spatial directions.
We study the momentum configuration $\pbf=(0,0,0)$ in the following.
The remaining dimensions are only the time direction and spinor space,
therefore for a given $L_s$ and the fixed spatial momentum configuration,
$\Delta^{(L_s)}$ is a matrix with dimension $4(T/a-1)$.
Figure \ref{fig:GWbreaking} shows the magnitude of
$\ln(||\Delta^{(L_s)}(x_0,y_0)||_{\rm spin})$,
where the norm is taken for the spinor space only.
By increasing $L_s$, the bulk symmetry breaking is reduced.
Finally at $L_s=24$, only boundary breaking effects remain
and they are localized exponentially near the time boundaries.
This is the expected behavior for overlap fermions \cite{Luscher:2006df}.
We conclude that the presence of the boundary term
causes chiral symmetry breaking, which decays exponentially away from
the time boundaries
for the effective four dimensional operator.

\section{Spectrum of free operator}
\label{sec:spectrum}
In this section, we investigate
the free spectrum of the DWF operator
to confirm universality at the tree level.
We set $T=L$ in this section.

\subsection{Spectrum of $\DWF^{\dag}\DWF$}
\begin{figure}[!t]
\begin{center}
  \scalebox{1.1}{\includegraphics{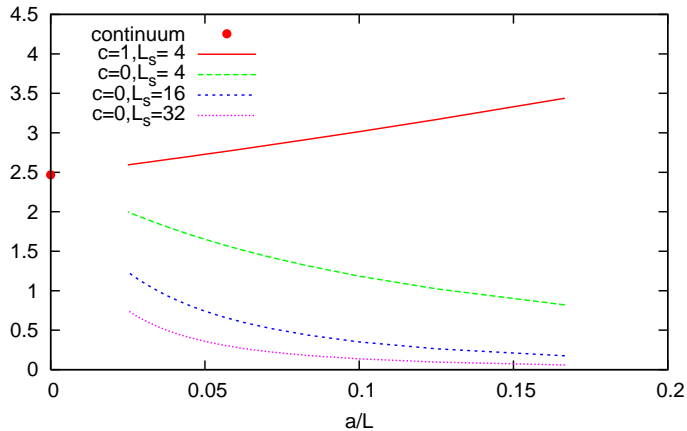}}
\caption{
The lowest eigenvalue of $L^2\DWF^{\dag}\DWF$ with $a m_5=1$.
Some combinations of parameters $c$ and $L_s$ are shown.
The continuum value is $\pi^2/4=2.467..$.
For $c=1$ case, since the $L_s$-dependence is so weak on on this scale,
we show only $L_s=4$ results as representative.
\label{fig:DWFEV}}
\end{center}
\end{figure}

\begin{figure}[!t]
\begin{center}
  \begin{tabular}{cc}
  \scalebox{1.1}{\includegraphics{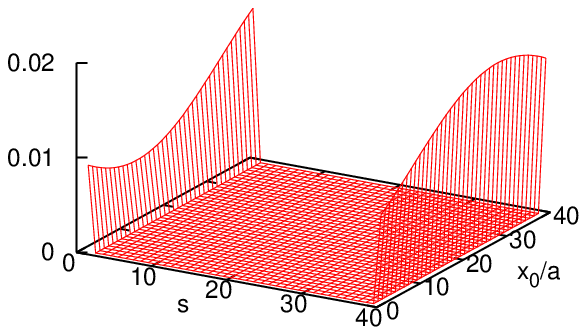}}&
  \scalebox{1.1}{\includegraphics{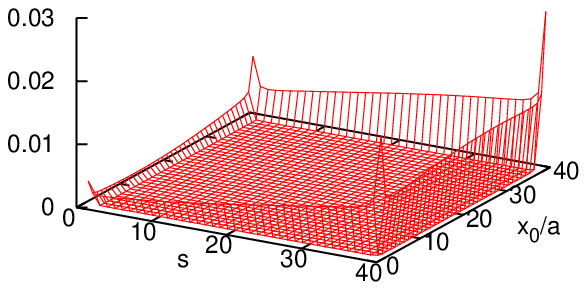}}\\
  \end{tabular}
\caption{$||\psi(x_0,s)||_{\rm spin}$
with zero spatial momentum and the parameters
$T/a=L_s=40$, $\theta=0$ and $am_5=1$.
The norm for the eigenvector is taken in the spinor space.
The left (right) panel is for $c=1$ ($c=0$).
\label{fig:WF}}
\end{center}
\end{figure}

To achieve better chiral symmetry,
the (physical) eigenmodes of the domain-wall operator should
be localized near the boundaries of the fifth direction
and propagate in the space-time directions.
This should also be true in the SF setup,
since the chiral symmetry is supposed to be maintained in the bulk.
To observe such phenomena, we numerically compute the lowest eigenmode
of the operator $\DWF^{\dag}\DWF$ with the trivial gauge configuration $U(x,\mu)=1$.
In the free case, 
we can perform the Fourier transformation for spatial directions,
and project out the momentum configuration $\pbf=(0,0,0)$.
Thus, remaining indexes of the vector space
are now the spinor, the time $x_0$ and the extra dimension $s$,
\be
\DWF^{\dag}\DWF\psi(x_0,s)=\lambda\psi(x_0,s),
\ee
where the spinor indexes are suppressed.
We set input parameters $am_5=1$ and $\theta=0$.
$\theta$ is the parameter which
controls the spatial boundary condition for fermion fields.
(For more details, we refer to \cite{Sint:1995ch}.)

We numerically compute the lowest eigenvalue and the corresponding
eigenfunction $||\psi(x_0,s)||_{\rm spin}$.
We examine not only for $c=1$ but also for $c=0$
to investigate the importance of the presence of the boundary operator.
The scaling behavior of the eigenvalue is shown in Figure \ref{fig:DWFEV}.
For $c=1$, $L_s$ dependence is too small to see on this scale.
The lowest eigenvalues converge to their continuum values properly,
therefore universality is confirmed.
Furthermore, the associated eigenfunction shows
nice localization behavior, namely
being localizing for the fifth direction and propagating for the time direction,
as shown in the left panel of Figure \ref{fig:WF}.
This shows that this mode is a physical one.

For $c=0$ in Figure \ref{fig:DWFEV},
although all $L_s=4,16,32$ results tend to converge to the continuum limit,
large $L_s$ results have a bending phenomenon
in small $a/L$ region and show no power decay in terms of $a/L$. 
This indicates that if one takes $L_s$ to infinite before taking $a/L=0$ limit,
the eigenvalue will likely converge to zero.
If this is so, the theory with $L_s=\infty$ does not
belong to a correct universality class.
Furthermore, the eigenfunction in the right panel in Figure \ref{fig:WF}
is localized on edges in the time-$s$ plane.
This is a typical unphysical mode.
On the other hand, interestingly for small $L_s$,
the scaling behavior is rather mild.
In the small $L_s$ case, the chiral symmetry breaking
of domain-wall fermions are rather similar
to that of the ordinary Wilson fermions.
As in the Wilson fermions case,
the bulk chiral symmetry breaking for DWFs due to finite $L_s$
plays some role in producing the correct continuum limit.
This is the reason why DWFs with smaller $L_s$ and no
boundary term $B$ can produce the continuum results.


The results shown in this subsection show
that the boundary term with $c\neq 0$ plays an important role
for the theory to be in the correct universality class.

\subsection{Spectrum of $\Dq^{\dag}\Dq$}

\begin{figure}[!t]
\begin{center}
  \begin{tabular}{cc}
  \scalebox{1.1}{\includegraphics{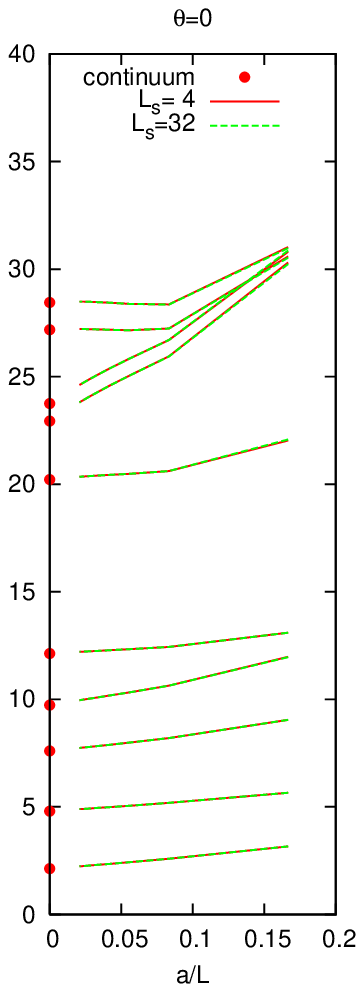}}&
  \scalebox{1.1}{\includegraphics{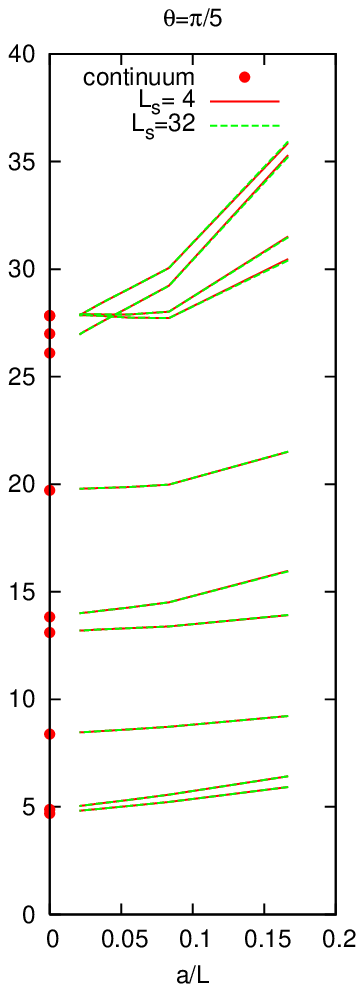}}\\
  \end{tabular}
\caption{The $a/L$ dependence of the lowest ten eigenvalues
of $L^2\Dq^{\dag}\Dq$ in the presence of the background gauge field.
The left (right) panel is for $\theta=0$ ($\theta=\pi/5$).
The parameters are set to $am_5=1$ and $c=1$.
The red points at $a/L=0$ are continuum values \cite{Sint:1995ch}.
\label{fig:lowest10}}
\end{center}
\end{figure}

Not all eigenmodes of $\DWF$
are physical ones and elimination of unphysical mode is not clear.
To extract physical modes only, let us study the eigenmodes
of $\Dq$, where unphysical modes are excluded.
The operator $\Dq$ is defined from $S_{\rm q}$ in eq.(\ref{eqn:Sq})
\be
\Dq S_{\rm q} (x,y)=a^{-4}\delta_{x,y}.
\ee

We numerically compute the lowest ten eigenvalues of
$\Dq^{\dag}\Dq$
with the parameter set $am_5=1$, $\theta=0,\pi/5$ and $c=1$
in the presence of the the background gauge field (choice A in
Ref.~\cite{Luscher:1993gh}).
The values obtained for $L/a=T/a=6,12,24$ are summarized in Table \ref{tab:MIN10}.
All tables are given in appendix \ref{sec:tables}
The scaling behavior of the eigenvalues
are shown in Figure \ref{fig:lowest10}.
Although we show two cases of $L_s$, namely $L_s=4$ and $L_s=32$,
it is hard to see the difference on this scale.
We observe that they converge to the continuum values given
in Ref~\cite{Sint:1995ch}.
This behavior persists for a variety of values of $c$,
$0.5 \le c\le 1.5$.
This confirms universality at the tree level.

\section{One-loop analysis of SF coupling}
\label{sec:oneloop}
To check further universality at the quantum level
and renormalizability, we perform the one-loop order calculation
of the SF coupling.

\subsection{Definition and results}

We compute the fermion contribution to
the SF coupling \cite{Sint:1995ch}
$p_{1,1}(L/a,L_s)$
(we set $L=T$ as usual)
at one-loop order
for massless domain-wall fermions.
The one-loop coefficient is given as
\be
p_{1,1}(L/a,L_s)
=
\left.
\frac{1}{k}
\frac{\pal}{\pal\eta}
\ln \det(\DWF/\PV)
\right|_{\eta=\nu=0},
\label{eqn:p11}
\ee
with a normalization (See \cite{Sint:1995ch} for details.)
\be
k=12(L/a)^2[\sin(\gamma)+\sin(2\gamma)],
\hspace{5mm}
\gamma=\frac{1}{3}\pi(a/L)^2.
\ee
The parameters $\eta$ and $\nu$ parameterize
the background gauge field \cite{Luscher:1993gh}.
In the actual calculation,
we expand the $\eta$ derivative
and use the fact that the determinant
is factorized for individual spatial momentum $\pbf$ and color sector $b$,
\bea
p_{1,1}(L/a,L_s)
&=&
\frac{1}{k}
{\rm Tr}
\left[
\DWF^{-1}\frac{\pal\DWF}{\pal\eta}
-\PV^{-1}\frac{\pal\PV}{\pal\eta}
\right]
\non\\
&=&
\frac{1}{k}
\sum_{\pbf}
\sum_{b=1}^{3}
{\rm tr}
\left[
(\DWF^b)^{-1}(\pbf)\frac{\pal\DWF^b(\pbf)}{\pal\eta}
\right.
\non\\&&
\left.
-(\PV^b)^{-1}(\pbf)\frac{\pal\PV^b(\pbf)}{\pal\eta}
\right].
\label{eqn:p11expand}
\eea
The trace
${\rm tr}$
concerns the spinor, the time indices and fifth coordinate only.
It is maybe worthwhile to note that for our definition of DWF,
\be
\frac{\pal\DWF}{\pal\eta}
=
\frac{\pal\PV}{\pal\eta}
\ee
holds since the mass term does not involve the gauge field.

We compute $p_{1,1}$ on the lattices of size $L/a=4,6,...,48$ and
$L_s=6,8,10,12,16$ with parameters $0.7 \le am_5 \le 1.3$
and $\theta=\pi/5$.
Subsets of the results are summarized in Table \ref{tab:p11}
for $am_5=1$, $L_s=6$ and $L/a=4,6,...,48$.
Separate contributions from DWF and PV are
also shown there.

\subsection{Coefficients of Symanzik's expansion}
\begin{figure}[!t]
\begin{center}
  \scalebox{1.3}{\includegraphics{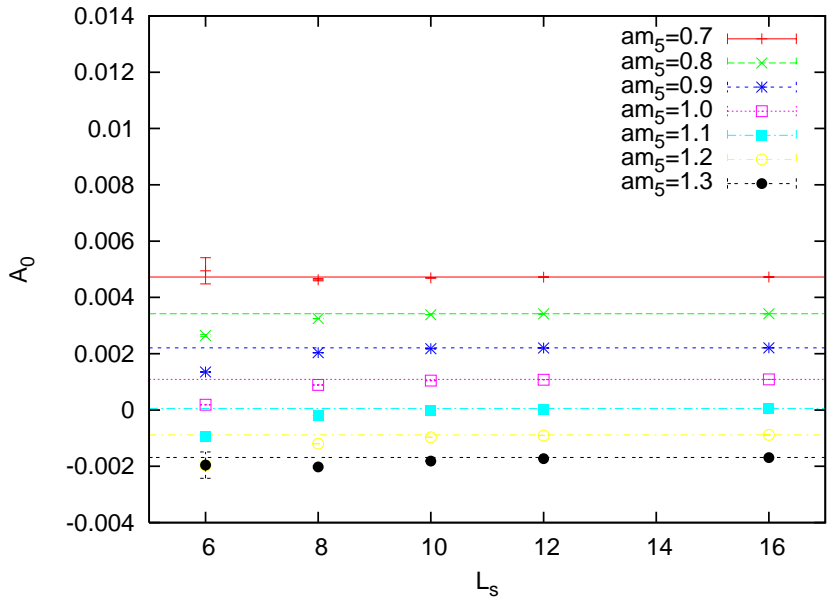}}
\caption{
$L_s$-dependence of $A_0$ for $0.7 \le am_5 \le 1.3$.
The horizontal lines show 
the values of $A_0$ in the infinity $L_s$ limit,
which are obtained by combining the results
of previous literature
\cite{Sint:1995ch,Aoki:2003uf}.
\label{fig:A0Ls}}
\end{center}
\end{figure}

From the Symanzik's analysis of the cutoff dependence
of Feynman diagrams on the lattice,
one expects that the one-loop coefficient has
an asymptotic expansion in terms of $a/L$
\begin{equation}
p_{1,1}(L/a,L_s)
=
\sum^{\infty}_{n=0}
(a/L)^n[A_n(L_s) + B_n(L_s) \ln (L/a)].
\label{eqn:p11symanzik}
\end{equation}
Note that the coefficients $A_n$ and $B_n$ ($n=0,1,2,...$)
depend on $L_s$.
We can reliably extract the first few coefficients
by making use of the method described in Ref.~\cite{Bode:1999sm}.

For the usual renormalization of the coupling constant,
$B_0$ at $L_s=\infty$ should be $2b_{0,1}$ where $b_{0,1}$ is
the fermion part of the one-loop coefficient
of the $\beta$-function for $N_{\rm f}$ flavors QCD,
\bea
b_0
&=&
b_{0,0}+N_{\rm f}b_{0,1},
\\
b_{0,0}
&=&
\frac{11}{(4\pi)^2},
\\
b_{0,1}
&=&
-\frac{2}{3}
\frac{1}{(4\pi)^2}.
\eea
We confirmed that
$B_0(L_s)$ for large $L_s$ (say $L_s=16$) converges to $2b_{0,1}=-0.008443...$
up to three significant digits
for the values of $am_5$ which we investigated.
When the tree-level O($a$) improvement is realized,
we expect that $B_1 = 0$ holds.
We check this to $10^{-3}$ for the same parameter region as before.
This shows that the formula
for the boundary coefficient in eq.(\ref{eqn:ctreeL})
works well to achieve the tree-level O($a$) improvement
to the precision considered here.
In the following analysis, we set exact values $B_0=2b_{0,1}=-1/(12 \pi^2)$
and $B_1=0$.

$A_0$ gives information about a ratio of $\Lambda$-parameters.
The obtained values of $A_0(L_s)$
as a function of $L_s$ are shown
in Figure \ref{fig:A0Ls}.
By combining the previous results
from Ref.~\cite{Sint:1995ch,Aoki:2003uf},
the values of $A_0$ at infinity $L_s$ can be obtained, and
are shown in Figure \ref{fig:A0Ls} as the horizontal lines.
We observe that our results at finite $L_s$
properly converge to the known results at infinity $L_s$.

To achieve one-loop O($a$) improvement,
we need to determine the coefficient
of the fermion part of the boundary
counter-term, $c_{\rm t}^{(1,1)}$ \cite{Sint:1995ch}
at one-loop order.
If one imposes an improvement condition
\cite{Sint:1995ch}, one finds that
\be
c_{\rm t}^{(1,1)}=A_1/2,
\label{eqn:ct11A1}
\ee
therefore we need the value of $A_1$.
The obtained values of $A_1$ are given in Table \ref{tab:A1}.
For future reference, we provide an interpolation formula for
$c_{\rm t}^{(1,1)}$ as a polynomial of $am_5$
for larger $L_s$, where value of $A_1$ is saturated,
\be
c_{\rm t}^{(1,1)}
=
0.00434+0.01102(am_5-1)-0.00858(am_5-1)^2,
\label{eqn:ct11}
\ee
for $0.7 \le am_5 \le 1.3$.

\section{Lattice artifacts of the step scaling function to one-loop order}
\label{sec:SSF}
\begin{figure}[!t]
 \begin{center}
  \begin{tabular}{cc}
  \resizebox{70mm}{!}{\includegraphics{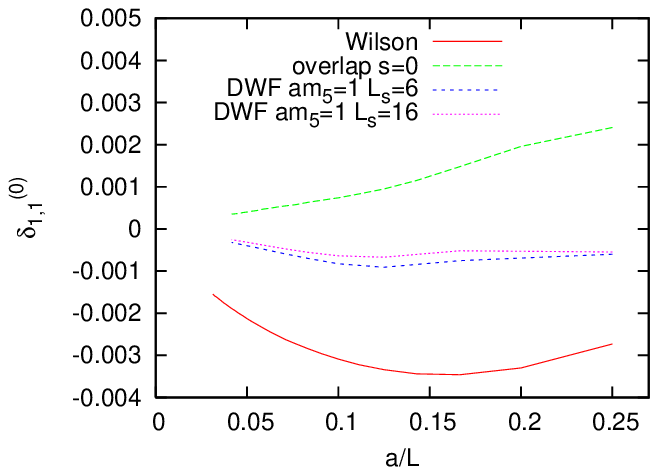}}
  &
  \resizebox{70mm}{!}{\includegraphics{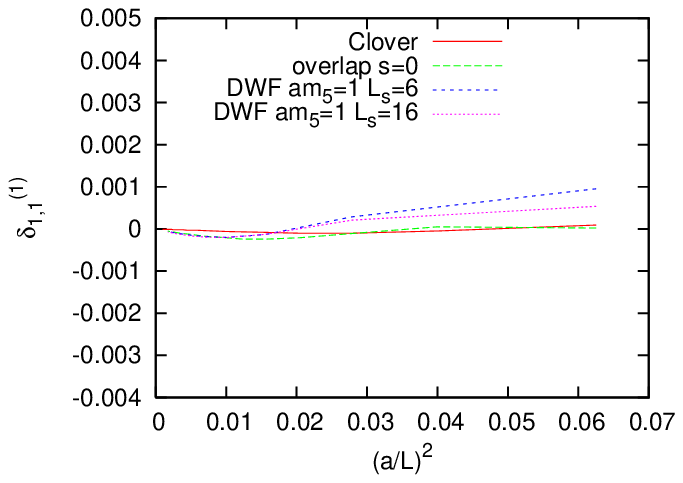}}
  \\
  \end{tabular}
  \caption{We show the relative deviation with the various actions for tree level
$O(a)$ improvement, $\delta_{1,1}^{(0)}$ (Left),
and one-loop $O(a)$ improvement, $\delta_{1,1}^{(1)}$ (Right),
as a function of $a/L$ and $(a/L)^2$ respectively.
Upper part is for $\theta=0$, and lower is for $\theta=\pi/5$.
For comparison, those of the Wilson type fermion
with $c_{\rm t}^{(1,1)}=0$ and the clover fermion with
$c_{\rm t}^{(1,1)}=0.019141$ \cite{Sint:1995ch}
are included in the plot of $\delta_{1,1}^{(0)}$ and
$\delta_{1,1}^{(1)}$ respectively.
  \label{fig:deviation}}
 \end{center}
\end{figure}

In this section, we investigate lattice artifacts of
the step scaling function (SSF) \cite{Luscher:1991wu}
$\sigma(2,u)$, which describes the evolution
of the running coupling $\bar{g}^2(L)=u$
under changes of scale $L$ by a factor $2$,
\be
\sigma(2,u)=\bar{g}^2(2L),
\hspace{5mm}
u=\bar{g}^2(L).
\ee
The lattice version of the step scaling function is
denoted by $\Sigma(2,u,a/L)$ which
contains lattice artifacts.
Such lattice artifacts
are described by the relative deviation
\be
\delta (u,a/L) 
\equiv
\frac{\Sigma(2,u,a/L) 
 - \sigma(2,u)}{\sigma(2,u)}.
\ee
By expanding the relative deviation
in terms of the coupling constant $u$, one obtains
\be
\delta (u,a/L) 
= 
\delta_1(a/L) u  
+ O(u^2),
\ee
where the one-loop deviation, $\delta_1(s,a/L)$,
may be decomposed into pure gauge and fermion part \cite{Sint:1995ch},
\be
\delta_1(a/L)
=
\delta_{1,0}(a/L)
+N_{\rm f}
\delta_{1,1}(a/L).
\ee
We are currently only interested in the fermion part.
We consider domain-wall fermions,
thus the fermion part of the one-loop deviation
$\delta_{1,1}(a/L,L_s)$
contains $L_s$ dependence.
In terms of the
one-loop coefficient of the SF coupling $p_{1,1}$,
the one-loop deviation is given by
\begin{equation}
\delta_{1,1}(a/L,L_s)
= 
 p_{1,1}(2L/a,L_s)
-p_{1,1}(L/a,L_s)
-2 b_{0,1} \ln(2).
 \label{eqn:deviation1}
\end{equation}
Depending on the value of the
boundary counter term $c_{\rm t}^{(1,1)}$,
we denote with $\delta^{(0)}_{1,1}$
the tree level O($a$) improved version
with $c_{\rm t}^{(1,1)}=0$,
and $\delta^{(1)}_{1,1}$
the  one-loop O($a$) improved one
for $c_{\rm t}^{(1,1)}$ in eq.(\ref{eqn:ct11A1}).


We show numerical results for the one-loop deviation in
Tables \ref{tab:deviation0} and \ref{tab:deviation1}
and the plots in Figure \ref{fig:deviation},
where we include those of the Wilson type fermions \cite{Sint:1995ch}
and overlap fermion \cite{Takeda:2007ga}
for comparison.
$L_s$-dependence of DWFs is small.
In the case of the clover action, $c_{\rm t}^{(1,1)}$ is set
to be the proper value to achieve one-loop $O(a)$ improvement,
and for Wilson fermions
it is set to $c_{\rm t}^{(1,1)}=0$.
We observe that the lattice artifacts
for domain-wall fermions are small for
tree level boundary O($a$) improvement case
compared with other fermions, while
they are large for one-loop boundary O($a$) improvement case.

\section{Conclusion and outlook}
\label{sec:conclusion}

In this paper, we provide a new formulation of domain-wall fermions
in the SF setup by following the universality argument of L\"uscher.
In contrast to the previous formulation by Taniguchi,
ours can deal with the boundary O($a$) improvement properly,
and there is no constraint on the number of flavors.
To check that our formulation works properly,
we investigate the spectrum and eigenmodes of the free operator,
and perform a one-loop analysis of the SF coupling constant.
Then we confirm universality at tree and the one-loop level
and observe that all results investigated
show the desired behaviors.

Before starting simulations,
the boundary improvement coefficient $c$
should be determined to one-loop order.
This involves calculations of the SF
correlators, $f_{\rm A}$, $f_{\rm P}$ etc. given in
appendix \ref{sec:operator}.
This could be done in a similar way to the case of the Wilson fermion.

As mentioned before, one of the most important
properties of the universality formulation
is that there are no restriction of the number of flavors.
By taking advantage of this property,
we may compute the renormalization factor
of $B_{\rm K}$ for $N_{\rm f}=3$ QCD.

\section*{Acknowledgments}
We would like to thank Sinya Aoki, Yasumichi Aoki,
Norman Christ, Michael Endres,
Taku Izubuchi, Changhoan Kim, Robert Mawhinney
and the members of the RBC Collaboration
for helpful discussions.
We are grateful to Stefan Sint for his hospitality
during our stay at Trinity College in Dublin,
where this work was initiated.
This work is supported by the U.S. Department of Energy
under Grant No. DE-FG02-92ER40699.

\appendix

\section{Fermion correlators}
\label{sec:operator}
In this appendix, I summarize
fermion correlators, the boundary fields and so on which are often used
in the SF setup.

\subsection{Boundary fields}
\label{subsec:boundaryfields}
As given in Ref.~\cite{Luscher:2006df},
the lattice version of the fermion boundary fields are defined
\bea
\zeta(\xbf)&=&U(x-a\hat{0},0)P_-q(x)|_{x_0=a},
\\
\bar\zeta(\xbf)&=&\bar{q}(x)P_+U(x-a\hat{0},0)^{-1}|_{x_0=a},
\\
\zeta^{\prime}(\xbf)&=&U(x,0)^{-1}P_+q(x)|_{x_0=T-a},
\\
\bar\zeta^{\prime}(\xbf)&=&\bar{q}(x)P_-U(x,0)|_{x_0=T-a}.
\eea
Here note that we use the physical quark fields
defined in eq.(\ref{eqn:q})
and (\ref{eqn:barq}).

\subsection{Propagators}
\label{subsec:propagators}
The propagators for the physical quark fields
and the boundary fields are given by
\bea
\left[q(x)\bar{q}(y)\right]_{\rm F} &=& S_{\rm q}(x,y),
\\
\left[q(x)\bar{\zeta}(\ybf)\right]_{\rm F} &=& S_{\rm q}(x,y)U(y-a\hat{0},0)^{-1}P_+|_{y_0=a},
\\
\left[q(x)\bar{\zeta}^\prime(\ybf)\right]_{\rm F}&=&S_{\rm q}(x,y)U(y,0)P_-|_{y_0=T-a},
\\
\left[\zeta(\xbf)\bar{q}(y)\right]_{\rm F}&=&P_-U(x-a\hat{0},0)S_{\rm q}(x,y)|_{x_0=a},
\\
\left[\zeta^\prime(\xbf)\bar{q}(y)\right]_{\rm F}&=&P_+U(x,0)^{-1}S_{\rm q}(x,y)|_{x_0=T-a},
\\
\left[\zeta(\xbf)\bar{\zeta}^\prime(\ybf)\right]_{\rm F}&=&P_-U(x-a\hat{0},0)S_{\rm q}(x,y)U(y,0)P_-|_{x_0=a,y_0=T-a},
\\
\left[\zeta^\prime(\xbf)\bar{\zeta}(\ybf)\right]_{\rm F}&=&P_+U(x,0)^{-1}S_{\rm q}(x,y)U(y-a\hat{0},0)^{-1}P_-|_{x_0=T-a,y_0=a}.
\eea

\subsection{Operators}
\label{subsec:operators}
We consider the degenerate quark mass case
and an extension of the flavor space is done in a trivial way.
In terms of the physical quark fields,
the local operators are defined as
\bea
A_\mu^a(x)&=&\bar{q}(x)\gamma_\mu\gamma_5\frac{1}{2}\tau^aq(x),
\\
P^a(x)&=&\bar{q}(x)\gamma_5\frac{1}{2}\tau^aq(x).
\eea
The conserved axial vector current is given by
\be
{\cal A}_\mu^a(x)=
\sum_{s=1}^{L_s}
{\rm sign}\left(s-\frac{L_s+1}{2}\right)j_\mu^a(x,s),
\ee
where
\bea
j_\mu^a(x,s)&=&
 \bar\psi(x+a\hat\mu,s)P_+^{(\mu)}U(x,\mu)^{-1}\frac{1}{2}\tau^a\psi(x,s)
\non\\
&&
-\bar\psi(x,s)P_-^{(\mu)}U(x,\mu)\frac{1}{2}\tau^a\psi(x+a\hat\mu,s),
\eea
with $P_\pm^{(\mu)}=(1\pm\gamma_\mu)/2$.

\subsection{Correlators}
\label{subsec:correlators}
The fermion correlators for the local operators in the SF
are given by
\bea
f_{\rm A}(x_0)&=&-a^6\sum_{a=1}^{N_{\rm f}}\sum_{\ybf,\zbf}\frac{1}{N_{\rm f}^2-1}
\langle A_0^a(x)\bar\zeta(\ybf)\gamma_5\frac{1}{2}\tau^a\zeta(\zbf)\rangle,
\\
f_{\rm P}(x_0)&=&-a^6\sum_{a=1}^{N_{\rm f}}\sum_{\ybf,\zbf}\frac{1}{N_{\rm f}^2-1}
\langle P^a(x)\bar\zeta(\ybf)\gamma_5\frac{1}{2}\tau^a\zeta(\zbf)\rangle,
\\
f_1&=&-\frac{a^{12}}{L^6}\sum_{a=1}^{N_{\rm f}}\sum_{\ubf,\vbf,\ybf,\zbf}\frac{1}{N_{\rm f}^2-1}
\langle\bar\zeta^{\prime}(\ubf)\gamma_5\frac{1}{2}\tau^a\zeta^{\prime}(\vbf)
\bar\zeta(\ybf)\gamma_5\frac{1}{2}\tau^a\zeta(\zbf)\rangle.
\eea
After Wick contraction, they become
\bea
f_{\rm A}(x_0)&=&a^6\sum_{\ybf,\zbf}\frac{1}{2}
\langle [\zeta(\zbf)\bar{q}(x)]_{\rm F}\gamma_0\gamma_5[q(x)\bar\zeta(\ybf)]\gamma_5\rangle,
\\
f_{\rm P}(x_0)&=&a^6\sum_{\ybf,\zbf}\frac{1}{2}
\langle [\zeta(\zbf)\bar{q}(x)]_{\rm F}\gamma_5[q(x)\bar\zeta(\ybf)]\gamma_5\rangle,
\\
f_1&=&\frac{a^{12}}{L^6}\sum_{\ubf,\vbf,\ybf,\zbf}\frac{1}{2}
\langle [\zeta(\zbf)\bar\zeta^{\prime}(\ubf)]_{\rm F}\gamma_5
[\zeta^{\prime}(\vbf)\bar\zeta(\ybf)]_{\rm F}\gamma_5
\rangle,
\eea
where the propagators are given in subsection \ref{subsec:propagators} and we have used
\be
\sum_{a=1}^{N_{\rm f}}\tr\left[\left(\frac{\tau^a}{2}\right)^2\right]
=\frac{N_{\rm f}^2-1}{2}.
\ee
These correlators
are the same as those of Wilson fermions
except that the propagators are replaced by those of the physical quark field $S_{\rm q}$.

For the conserved axial vector current,
a correlator is given by
\be
f_{\cal A}(x_0)
=
-a^6\sum_{a=1}^{N_{\rm f}}\sum_{\ybf,\zbf}\frac{1}{N_{\rm f}^2-1}
\langle {\cal A}_0^a(x)\bar\zeta(\ybf)\gamma_5\frac{1}{2}\tau^a\zeta(\zbf)\rangle.
\ee
As an example, at tree level,
this can be expressed in terms of the propagator for domain-wall fermions as,
\bea
f_{\cal A}(x_0)|_{U=1}
&=&
\sum_{s=1}^{L_s}
{\rm sign}(s-\frac{L_s+1}{2})\frac{1}{2}{\rm Tr}
\left[\frac{}{}
\right.
\non\\&&
-P_+S_{\rm DWF}(x,y;s,1)\gamma_0P_LS_{\rm DWF}(y,x+a\hat{0};1,s)
\non\\&&
+P_-S_{\rm DWF}(x+a\hat{0},y;s,1)\gamma_0P_LS_{\rm DWF}(y,x;1,s)
\non\\&&
+P_+S_{\rm DWF}(x,y;s,1)P_RS_{\rm DWF}(y,x+a\hat{0};L_s,s)
\non\\&&
-P_-S_{\rm DWF}(x+a\hat{0},y;s,1)P_RS_{\rm DWF}(y,x;L_s,s)
\non\\&&
-P_+S_{\rm DWF}(x,y;s,L_s)P_LS_{\rm DWF}(y,x+a\hat{0};1,s)
\non\\&&
+P_-S_{\rm DWF}(x+a\hat{0},y;s,L_s)P_LS_{\rm DWF}(y,x;1,s)
\non\\&&
+P_+S_{\rm DWF}(x,y;s,L_s)\gamma_0P_RS_{\rm DWF}(y,x+a\hat{0};L_s,s)
\non\\&&
\left.
-P_-S_{\rm DWF}(x+a\hat{0},y;s,L_s)\gamma_0P_RS_{\rm DWF}(y,x;L_s,s)
\frac{}{}
\right].
\eea

\section{Tables of numerical results}
\label{sec:tables}

\begin{table}[h]
{\small
 \begin{center} 
  \begin{tabular}{|c|rr|rr|rr|c|c|}
  \hline \hline
  \multicolumn{9}{|c|}{$\theta=0$} \\ \hline
  & \multicolumn{2}{c|}{$L/a=6$}
  & \multicolumn{2}{c|}{$L/a=12$}
  & \multicolumn{2}{c|}{$L/a=24$} & & \\
  $n$ & $L_s=4$ & $L_s=32$ & $L_s=4$ & $L_s=32$ & $L_s=4$ & $L_s=32$& $b$   & $d$ \\ \hline
  1&       3.161141&       3.160760&       2.591269&       2.591267&       2.350053&       2.350053&  2&  2\\
  2&       5.658392&       5.658925&       5.191148&       5.191161&       4.990293&       4.990293&  2&  2\\
  3&       9.050173&       9.045312&       8.196497&       8.196424&       7.888424&       7.888424&  3&  2\\
  4&      11.981137&      11.963010&      10.635061&      10.634737&      10.177881&      10.177878&  1&  2\\
  5&      13.098016&      13.101850&      12.434736&      12.434839&      12.281233&      12.281235&  3&  2\\
  6&      22.037107&      22.078353&      20.612026&      20.613136&      20.436083&      20.436098&  1&  2\\
  7&      30.304338&      30.232012&      25.944369&      25.942289&      24.578258&      24.578235&  2&  2\\
  8&      30.810696&      30.865401&      26.708727&      26.709232&      25.378691&      25.378700&  2&  2\\
  9&      30.585955&      30.530965&      27.241272&      27.239584&      27.184415&      27.184395&  1&  6\\
 10&      31.026393&      30.978110&      28.357563&      28.355829&      28.438888&      28.438866&  3&  6\\
  \hline \hline
  \multicolumn{9}{|c|}{$\theta=\pi/5$} \\ \hline
  & \multicolumn{2}{c|}{$L/a=6$}
  & \multicolumn{2}{c|}{$L/a=12$}
  & \multicolumn{2}{c|}{$L/a=24$} & & \\
  $n$ & $L_s=4$ & $L_s=32$ & $L_s=4$ & $L_s=32$ & $L_s=4$ & $L_s=32$& $b$   & $d$ \\ \hline
  1&       5.924559&       5.922886&       5.232916&       5.232896&       4.952553&       4.952553&  2&  2\\
  2&       6.428276&       6.423570&       5.566868&       5.566810&       5.214696&       5.214695&  1&  2\\
  3&       9.221621&       9.223322&       8.721989&       8.722031&       8.548948&       8.548949&  2&  2\\
  4&      13.912656&      13.926143&      13.392223&      13.392552&      13.267533&      13.267537&  1&  2\\
  5&      15.970852&      15.951467&      14.513964&      14.513548&      14.162258&      14.162253&  3&  2\\
  6&      21.508691&      21.520321&      19.983404&      19.983750&      19.838031&      19.838036&  3&  2\\
  7&      35.296282&      35.214384&      29.231042&      29.228128&      27.743584&      27.743548&  2&  2\\
  8&      35.861910&      35.932144&      30.057783&      30.058382&      28.612335&      28.612346&  2&  2\\
  9&      30.468452&      30.406862&      27.727188&      27.725168&      27.806412&      27.806386&  1&  6\\
 10&      31.520474&      31.475867&      28.021244&      28.019681&      27.896264&      27.896245&  3&  6\\
  \hline \hline
  \end{tabular}
 \end{center}
 \caption{The lowest ten eigenvalues
of the Hermitian operator $L^2 \Dq^{\dag}\Dq$ for $L_s=4,32$.
Upper (Lower) panel is for $\theta=0$ ($\theta=\pi/5$).
$b$ represents the color sector, and $d$ is for degeneracy for
one flavor.
}
\label{tab:MIN10}
}
\end{table}

\begin{table}[t]
{\small
 \begin{center} 
  \begin{tabular}{|c|c|c|c|}
  \hline \hline
    $L/a$ & $p_{1,1}(L/a,6)$ & DWF contribution & PV contribution \\
\hline
  4&  -0.0090558230&  -0.0463742376&  -0.0373184146\\
  6&  -0.0128831139&  -0.0614487481&  -0.0485656342\\
  8&  -0.0155107254&  -0.0717039592&  -0.0561932338\\
 10&  -0.0176883193&  -0.0786362064&  -0.0609478871\\
 12&  -0.0194881626&  -0.0835634805&  -0.0640753179\\
 14&  -0.0209908578&  -0.0872753290&  -0.0662844712\\
 16&  -0.0222708990&  -0.0902037775&  -0.0679328785\\
 18&  -0.0233830732&  -0.0925954634&  -0.0692123901\\
 20&  -0.0243655725&  -0.0946007679&  -0.0702351955\\
 22&  -0.0252452627&  -0.0963170183&  -0.0710717556\\
 24&  -0.0260415528&  -0.0978103244&  -0.0717687716\\
 26&  -0.0267688671&  -0.0991273624&  -0.0723584952\\
 28&  -0.0274382013&  -0.1003021397&  -0.0728639383\\
 30&  -0.0280581232&  -0.1013600906&  -0.0733019674\\
 32&  -0.0286354356&  -0.1023206640&  -0.0736852285\\
 34&  -0.0291756284&  -0.1031990186&  -0.0740233902\\
 36&  -0.0296831955&  -0.1040071669&  -0.0743239714\\
 38&  -0.0301618611&  -0.1047547685&  -0.0745929075\\
 40&  -0.0306147460&  -0.1054496923&  -0.0748349463\\
 42&  -0.0310444913&  -0.1060984224&  -0.0750539311\\
 44&  -0.0314533516&  -0.1067063577&  -0.0752530061\\
 46&  -0.0318432669&  -0.1072780356&  -0.0754347687\\
 48&  -0.0322159188&  -0.1078173020&  -0.0756013832\\
  \hline \hline
  \end{tabular}
 \end{center}
 \caption{The one-loop coefficient of the SF coupling
$p_{1,1}(L/a,L_s)$ with $L_s=6$, $am_5=1$ and $\theta=\pi/5$.
In eq.(\ref{eqn:p11expand}), there are two sources
of contributions: DWF and PV, which are shown
separately in the table.
}
\label{tab:p11}
}
\end{table}

\begin{table}[b]
{\small
 \begin{center} 
  \begin{tabular}{|c|c|c|c|c|c|c|c|}
  \hline \hline
    $L_s$ $\diagdown$ $am_5$ & 0.7 & 0.8 & 0.9 & 1.0 & 1.1 & 1.2 & 1.3 \\
\hline
 6&   -      &   -     &0.0102(8)&0.0125(9)&0.0145(7)&    -    &-\\
 8&   -      &0.0047(4)&0.0074(9)&0.0097(9)&0.0119(9)&0.0135(4)&-\\
10&0.0004(2) &0.0040(9)&0.0066(9)&0.0090(9)&0.0111(9)&0.0129(9)&0.0135(2)\\
12&0.0007(9) &0.0037(9)&0.0064(9)&0.0088(9)&0.0108(9)&0.0126(9)&0.0139(8)\\
16&0.0006(10)&0.0036(9)&0.0063(9)&0.0087(9)&0.0107(9)&0.0125(9)&0.0137(10)\\
  \hline \hline
  \end{tabular}
 \end{center}
 \caption{The value of $A_1$ for $L_s=6,8,10,12,16$ and
$0.7\le am_5\le 1.3$.
}
\label{tab:A1}
}
\end{table}

\begin{table}[t]
{\small
 \begin{center} 
  \begin{tabular}{|c|c|c|c|c|c|c|c|c|c|c|}
  \hline \hline
$L/a$ $\diagdown$ $L_s$ & 6&8&10&12&16\\
\hline
 4&-0.000602&-0.000518&-0.000527&-0.000538&-0.000545\\
 6&-0.000753&-0.000562&-0.000522&-0.000516&-0.000517\\
 8&-0.000908&-0.000737&-0.000688&-0.000674&-0.000669\\
10&-0.000825&-0.000695&-0.000655&-0.000643&-0.000637\\
12&-0.000701&-0.000599&-0.000569&-0.000560&-0.000555\\
14&-0.000595&-0.000511&-0.000487&-0.000480&-0.000476\\
16&-0.000512&-0.000439&-0.000419&-0.000413&-0.000411\\
18&-0.000448&-0.000383&-0.000365&-0.000360&-0.000358\\
20&-0.000397&-0.000338&-0.000322&-0.000318&-0.000316\\
22&-0.000356&-0.000302&-0.000288&-0.000284&-0.000282\\
24&-0.000322&-0.000273&-0.000259&-0.000256&-0.000254\\
  \hline \hline
  \end{tabular}
 \end{center}
 \caption{The relative deviation $\delta_{1,1}^{(0)}$ with
$am_5=1$ and $\theta=\pi/5$ for tree level boundary O($a$) improvement.
}
\label{tab:deviation0}
}
\end{table}

\begin{table}[b]
{\small
 \begin{center} 
  \begin{tabular}{|c|c|c|c|c|c|c|c|c|c|c|}
  \hline \hline
$L/a$ $\diagdown$ $L_s$ & 6&8&10&12&16\\
\hline
 4& 0.000955& 0.000701& 0.000597& 0.000558& 0.000539\\
 6& 0.000286& 0.000251& 0.000227& 0.000214& 0.000206\\
 8&-0.000129&-0.000128&-0.000126&-0.000127&-0.000127\\
10&-0.000202&-0.000207&-0.000206&-0.000204&-0.000203\\
12&-0.000182&-0.000193&-0.000195&-0.000194&-0.000194\\
14&-0.000150&-0.000162&-0.000166&-0.000166&-0.000166\\
16&-0.000123&-0.000134&-0.000138&-0.000139&-0.000140\\
18&-0.000101&-0.000112&-0.000115&-0.000117&-0.000117\\
20&-0.000085&-0.000094&-0.000098&-0.000099&-0.000099\\
22&-0.000072&-0.000080&-0.000083&-0.000084&-0.000085\\
24&-0.000062&-0.000069&-0.000072&-0.000073&-0.000074\\
  \hline \hline
  \end{tabular}
 \end{center}
 \caption{The relative deviation $\delta_{1,1}^{(1)}$ with
$am_5=1$ and $\theta=\pi/5$ for one-loop level boundary O($a$) improvement.
}
\label{tab:deviation1}
}
\end{table}



\providecommand{\href}[2]{#2}\begingroup\raggedright\endgroup

\end{document}